\documentclass[twocolumn]{aastex63}
\usepackage{amsmath} 
\usepackage{hyperref}
\usepackage{bm}
\usepackage{multirow}
\usepackage{gensymb}
\usepackage{enumitem}
\usepackage{tabularx}
\usepackage[flushleft]{threeparttable}
\usepackage[]{color}

\usepackage{natbib}
\usepackage{float}

\usepackage{amsfonts,amsmath,amssymb}
\usepackage{graphicx}
\usepackage{color}
\usepackage{xcolor}
\usepackage{natbib}
\usepackage{url}

\usepackage{breqn}
\usepackage{amsmath}
\usepackage{ulem}
\usepackage{soul} 

\newcommand{\xstar}{\textsc{\scriptsize XSTAR}}

\newcommand{\E}[1]{\left\langle#1\right\rangle}

\def\hep0{{\rm HEP_{0}}}

\def\<{\,\langle\langle}
\def\>{\,\rangle\rangle}

\defcitealias{Dannen20}{D20}
\defcitealias{Waters21}{WPD21}
\defcitealias{Begelman83}{BMS83}
\defcitealias{CAK}{CAK}
\defcitealias{Dannen24}{D24}

\def\CIVdbl{{\rm C~}\kern 0.1em{\sc iv}~$\lambda\lambda 1548, 1550$} 
\def\OVIIIi{\hbox{{\rm O}\kern 0.1em{\sc viii}}}
\def\SiXIVi{\hbox{{\rm Si}\kern 0.1em{\sc viii}}}
\def\OVIII{\hbox{{\rm O}\kern 0.1em{\sc viii}~{\rm Ly}$\alpha$}}
\def\SiXIV{\hbox{{\rm Si}\kern 0.1em{\sc viii}~{\rm Ly}$\beta$}}
\def\FeXXV{\hbox{{\rm Fe}\kern 0.1em{\sc xxv}}}
\def\FeXXVI{\hbox{{\rm Fe}\kern 0.1em{\sc xxvi}}}
\def\FeXXVK{\hbox{{\rm Fe}\kern 0.1em{\sc xxv}~{\rm K}$\alpha$}}
\def\FeXXVIK{\hbox{{\rm Fe}\kern 0.1em{\sc xxvi}~{\rm K}$\alpha$}}

\newcommand{\beq}{\begin{equation}}
\newcommand{\seq}{\end{equation}}

\renewcommand{\v}[1]{\ensuremath{\mathbf{#1}}} 
\newcommand{\abs}[1]{\left| #1 \right|} 
\renewcommand{\d}[2]{\frac{d #1}{d #2}} 

\newcommand{\f}{\frac}  
\newcommand{\e}[1]{\exp\left( #1 \right)} 

\usepackage{letltxmacro,amsmath}
\LetLtxMacro{\originaleqref}{\eqref}
\renewcommand{\eqref}{eq.~\originaleqref}

\submitjournal{ApJ}
\shorttitle{}
\shortauthors{Key et al.}
\begin{document}
\title{Oscillatory Line-Driven Winds: The Role of Atmospheric Stratification}

\correspondingauthor{Joshua A. Key}
\author[0000-0002-0942-637X]{Joshua A. Key}
\email{keyj5@unlv.nevada.edu}
\affiliation{Department of Physics \& Astronomy \\
University of Nevada, Las Vegas \\
4505 S. Maryland Pkwy \\
Las Vegas, NV, 89154-4002, USA}

\author[0000-0002-6336-5125]{Daniel Proga}
\affiliation{Department of Physics \& Astronomy \\
University of Nevada, Las Vegas \\
4505 S. Maryland Pkwy \\
Las Vegas, NV, 89154-4002, USA}

\author[0000-0002-5160-8716]{Randall Dannen}
\affiliation{Department of Physics \& Astronomy \\
University of Nevada, Las Vegas \\
4505 S. Maryland Pkwy \\
Las Vegas, NV, 89154-4002, USA}

\author[0000-0002-6336-5125]{Sterling Vivier}
\affiliation{Department of Physics \& Astronomy \\
University of Nevada, Las Vegas \\
4505 S. Maryland Pkwy \\
Las Vegas, NV, 89154-4002, USA}

\author[0000-0002-5205-9472]{Tim Waters}
\affiliation{Theoretical Division, Los Alamos National Laboratory}

\begin{abstract}

Time-dependent numerical studies of line-driven winds using the Sobolev approximation
have a history spanning over three decades.
In many of these studies, the wind solutions display notorious oscillations.
Two clues suggest the oscillations originate at the wind base:
(i) simulations reach a steady state without oscillations when the base density is sufficiently low and
(ii) the oscillation dominant frequency is comparable to the Lamb cut-off frequency, $\omega_c$, of acoustic waves propagating in a stratified hydrostatic atmosphere.
Recently, Dannen et al. observed another clue: when the line force
significantly weakens due to ionization, the winds become increasingly sensitive
to the self-excited oscillations. Here, we present a set of simulations
and \textit{perturbation analyses} that further elucidate the source and characteristics of oscillations.
We found that the line force adds wave energy and amplifies perturbations
with frequencies near $\omega_c$.
This selective amplification results from the coupling between the natural tendency
of velocity perturbations to grow in a stratified atmosphere
and from the line force dependence on the velocity gradient,
per the Castor-Abbott-Klein line-driven wind theory.
We also found that the variability stems from self-excitation that occurs
in the exponential atmosphere due to the non-linearity introduced
by the absolute value of the velocity gradient in the line force prescription.
We conclude that self-consistently calculating ionization is insufficient
for modeling the dynamics in the subsonic atmosphere. Instead, future wind
or unified models should relax the Sobolev approximation,
or model the radiative transfer to properly capture
the resulting radiation-induced instabilities and dynamics at the wind base.

\end{abstract}

\section{Introduction}\label{sec:intro}

Radiation-hydrodynamical problems — such as radiation-driven astronomical outflows — are challenging as they
are non-linear and require proper coupling of radiation, fluid dynamics, and ionization physics. 
The gas opacity, metallicity, and temperature will strongly affect this coupling. 
In a pivotal work by Castor, Abbott, and Klein, they showed that the bound-bound transitions of mildly ionized metals significantly enhance the opacity in the atmospheres of OB stars (\citeauthor{CAK} \citeyear{CAK}; hereafter \citetalias{CAK}), making substantial mass loss inevitable in such stars \cite[see also,][]{LS1970}.

In their line-driven wind theory, \citetalias{CAK} applied the Sobolev approximation, which allows the intrinsically non-local problem of radiative transfer in spectral lines to be treated as a local problem \citep{sobolev1960brightness}.
Consequently, the equations that describe the behavior
of the line-driven winds can be simplified. \citetalias{CAK} showed that the time-independent wind equations 
have a unique steady-state transonic solution. 
Formally the \citetalias{CAK} wind solution is invalid
in the subsonic region, {\it nearly} in hydrostatic equilibrium (HSE), because the Sobolev approximation relies on the flow velocity of the gas to exceed the thermal speed of the ions.
However, one can argue that this formal violation of the Sobolev approximation
is insignificant as the subsonic region is dense and the line force
is negligible. Therefore, the line force contributes little to the net force and the primary support against gravity comes from a combination of gas pressure and radiation pressure from continuum opacity sources.
In fact, the CAK wind solutions have many properties similar to those observed in stellar winds 
\citep[e.g., see a review in][and references therein]{LC}. 

In Time-dependent numerical simulations using the \citetalias{CAK} formalism, many simulations reach the same steady-state solution as the time-independent \citetalias{CAK} solution. Some simulations show a periodic oscillatory behavior especially simulations where the wind base was assumed to be dense, hence the base is in near HSE (see below for some examples).

These periodic oscillations have a frequency
approximately the same as the Lamb cut-off frequency \citep{Lamb1909theory}
\begin{equation}
    \label{eq:AcF}
    \omega_c=\frac{a}{2H},
\end{equation}
of acoustic waves with the sound speed, $a$ in an atmosphere in HSE with the density scale height, $H$. 
In early work, \cite{blondin1990hydrodynamic} observed the perturbing influence of Lamb oscillations on the outer wind structure of their mass transfer simulations. Others noted that line-driving solutions could exhibit oscillations if density at the inner boundary, $\rho_0$, is relatively high \citep[e.g.,][]{Owocki94, Proga98} or if the wind is relatively cold \cite{Proga99b}. 
\citet{feldmeier2001hydrodynamics} speculated that this "Lamb ringing" could produce 
"coherent trains of pronounced shells in hot star winds" but he also noted that the X-ray observations from hot star winds 
are more consistent with random variability instead. As such, many groups, regarding the oscillations as physically irrelevant, adopted the practice of setting the density low enough to remove the "long-lived oscillations" but high enough to capture the transonic region of the wind \citep[e.g.,][]{Proga99b, sundqvist2013clumping}. This practice continues 
even in the recent "unified" models that include both the stellar atmosphere and the wind. 
In the HSE region of 1D unified models, the density must be set appropriately; otherwise, Lamb oscillations 
will arise \citep[Jon Sunqvist, priv. comm.]{poniatowski2021dynamically, poniatowski2022method}. 

The success of the CAK formalism and its tractability in 2D and 3D time-dependent simulations, led to its application in  accretion disk systems such as cataclysmic variables \cite[see][for a review]{P05} 
and some active galactic nuclei \cite[AGNs; e.g.,][]{Mushotzky72, Arav94, MCGV, PSK, P07}. AGNs emit radiation over a broad energy range with a relatively large number of ionizing photons. 
Therefore, in some parts of their atmospheres, the ionization could be too high
to sustain mildly ionized metals \citep[and references therein]{Sim2010, Higginbottom14}.
To address this concern,  \cite{Dannen20} and
\citeauthor{Dannen24} (\citeyear{Dannen24}; hereafter \citetalias{Dannen24})
developed self-consistent wind models where hydrodynamical (`hydro' for short) calculations 
of the gas dynamics are coupled with ionization balance calculations \citep[see also][]{Dyda17, Dannen19}.
\citetalias{Dannen24} surveyed the parameter space of the line force resulting from blackbody (BB) 
spectral energy distributions (SEDs) with temperatures ranging from $\sim 10^4$ to 10$^6$~K and found that line driving remains efficient up to the temperature $\approx 4 \times10^5$~K. 
In addition, they ran a set of exploratory 1-D spherical wind simulations 
and found that the time-dependent solutions reached a steady state only in a subset of cases.
\citetalias{Dannen24} discovered that for the BB temperature larger than $8 \times10^5$~K, 
the line driven winds appeared intrinsically variable.

To explain the variability in their self-consistent simulations, 
\citetalias{Dannen24} examined a few isothermal cases using 
a modified CAK method (mCAK; hereafter) introduced by \cite{Abbott82} 
to study the effect of ionization on the line force. 
Specifically, \citetalias{Dannen24} approximated the line force--ionization relation 
with a power-law scaling $\xi^{-\delta}$, where $\delta$ is the parameter introduced 
by \cite{Abbott82} and $\xi$ is the ionization parameter (see below for more details).
They found that even for these simplified isothermal cases, which exclude thermal instability, 
the wind can be subject to self-excited oscillations if the line force weakens significantly with 
increasing ionization (e.g., $\delta~=~0.3$). Therefore, they attributed the variability 
in the solutions to the weakening of the line force instead 
of the thermal instability that they have observed in previous simulations 
with an AGN type of SED \citepalias{Dannen20}.

Moreover, they observed that the time-behavior of the solution depends on the so-called
hydrodynamic escape parameter at the inner radius, $r_0$, that is,
\begin{equation} \label{eq:hep}
\hep0 \equiv \frac{G {\rm M} (1 - \Gamma)}{r_0 a^{2}},
\end{equation}
where $\rm M$ is the central mass, 
and $\Gamma$ is the luminosity in units of the Eddington luminosity $L_{Edd}$. As well as, the density at the inner boundary $\rho_0$ and the ionization parameter $\delta$.
Specifically, the wind is time-independent when the atmosphere is hot or warm, $\hep0 < 50$ and $\delta<0.23$. However,
when the wind base is cold, $\hep0 \gtrsim 50$, the wind could be intrinsically variable
\citepalias[see Table~2 in ][]{Dannen24}. 

\citetalias{Dannen24} provided some clues regarding the nature of the variable solutions. 
Still, they did not identify the root cause of the variability.
They left this task for future investigation. This paper presents our follow-up analysis 
and findings regarding the persistent periodic behavior of isothermal line-driven winds. 

Below, we demonstrate that the line force produces self-excited perturbations 
near the wind base that evolve into radiative-acoustic waves in the wind. 
The line force preferentially amplifies low-frequency components of the self-excited perturbations, making the dominant frequency of order $\omega_c$. 
This selective amplification results from the coupling between the natural tendency 
of velocity perturbations to grow in a stratified atmosphere 
\citep[see, e.g.,][]{clarke2007principles} and the dependence of the line force 
on the velocity gradient, per the \citetalias{CAK} line-driven wind theory.
We also provide our current understanding of the process leading to self-excitation of perturbations
that operates at the deepest layers of the atmosphere where the flow velocity is tiny and CAK line force parametrization introduces a non-linear effect.

The paper is organized as follows. In \S~\ref{sec:method},
we describe our methods. 
In \S~\ref{sec:results}, we present the results of our analysis 
of the hydro calculations and our solutions of the wave equation.
In \S~\ref{sec:Disc&Conc}, we discuss our results and conclude.

\section{Methods}\label{sec:method}

To take a step forward in reducing the number of free parameters, \citetalias{Dannen24} 
used the photoionization code 
\xstar\footnote{\url{https://heasarc.nasa.gov/lheasoft/xstar/xstar.html}}~\citep{KB01} 
to self-consistently compute the radiative heating and cooling rates 
and the radiation force due to spectral lines on the gas due for various SEDs. They parameterized the heating and cooling rates with gas temperature, $T$, 
and ionization parameter 
\begin{equation}
    \label{eq:Xiparam}
    \xi~\equiv~(4\pi)^2\frac{J_X}{n_{\rm H}},
\end{equation}
where 
$J_X$ is the mean intensity integrated over the 0.1-1000 Ry range  and $n_{\rm H}$ is the number density of hydrogen nucleons.
\citetalias{Dannen24} computed and parameterized the line force with the aforementioned ionization parameter and optical depth parameter, $t=\sigma_{e} \rho \lambda_{Sob}$,
where $\sigma_{e}$ is the mass scattering coefficients for free electrons and 
$\lambda_{Sob}=~v_{th}/|v'|$ where $v_{th}$ is the gas thermal velocity and 
$v'=\partial v/\partial r$ is the velocity gradient of the gas in the radial direction.  
They found that for a given mass and SED, 
the efficiency of line driving and the wind solution depends on
three dimensionless parameters: $\hep0$, $\Gamma$, and $\Xi_0$.
The latter parameter is the pressure ionization parameter defined 
at the wind base 
\begin{equation}
    \label{eq:Xi_0}
    \Xi_0~\equiv~\xi_0/(4\pi c k_B T_0).
\end{equation}
The parameter $\hep0$ sets the strength of the thermal driving in the solution, 
while the parameters $\Gamma$ and $\Xi_0$ set the strength of the line driving 
by determining the line optical depth parameter, $t_0$ 
at the wind base (see eq.~B8 in \citetalias{Dannen24}). 
Typically, the number of driving lines increases with decreasing $\Xi_0$. 
However, for relatively small $\Xi_0$, the optical depth could become 
too large for the line force to operate despite the presence of many driving lines.

Most of the data analyzed in this paper come from simulations carried 
out by \citetalias{Dannen24}. However, to pin down
the mechanism responsible for variability,  
we also performed additional simulations using the same methodology 
as the original ones. Therefore, we will provide only a summary of the numerical method 
below \citepalias[for more details, refer to][]{Dannen24}.

\subsection{Numerical Simulations}
We use the magnetohydrodynamic code Athena++ \citep{Stone20} to solve
the equations of hydrodynamics for 1-D spherical winds driven by a radial radiation field 
\begin{equation}
  \label{eq:ContEq}
  \frac{\partial \rho}{\partial t} + \nabla \cdot (\rho \mathbf{v})~=~0,
\end{equation}
\begin{equation}
  \label{eq:MomEq}
  \frac{\partial (\rho \mathbf{v})}{\partial t} + \nabla \cdot (\rho \mathbf{v} \mathbf{v} + \mathbf{P})~=~-\rho \nabla\Phi_{\text{eff}} + \mathbf{g}_{l},
\end{equation}
and
\begin{equation}
  \label{eq:EnEq}
  \frac{\partial E}{\partial t} + \nabla \cdot \left[(E + p)\mathbf{v}\right]~=~-\rho\mathbf{v} \cdot \nabla\Phi_{\text{eff}} + \mathbf{v} \cdot \mathbf{g}_{l},
\end{equation}
where $\rho$, $\mathbf{P}$, $\mathbf{v}$, and $E$ are the gas density, pressure, velocity, 
and energy density, respectively. In addition,  $\mathbf{P}~=~p\mathbf{I}$ is the pressure tensor
where $p$ is the scalar pressure and $\mathbf{I}$ is the unit tensor. 
Moreover, $E~=~\rho\mathcal{E} + 1/2\rho|\mathbf{v}|^{2}$ is the energy density 
while $\Phi_{\text{eff}}~=~-GM(1-\Gamma)/r$ is the effective gravitational potential.

We compute the line force, $\mathbf{g}_{l}$ using the method introduced 
by \citetalias{CAK}
\begin{equation}
  \label{eq:Frad}
    \mathbf{g}_{l}~=~M(t,\xi) g_{\text{es}}\mathbf{\hat{r}},
\end{equation}
where $M(t,\xi)$ is the force multiplier and $g_{es}$ is the radiation force due to electron scattering. 
We approximate the force multiplier dependence on the optical depth, $t$, and ionization parameter, $\xi$,
using power laws:
\begin{equation}
  \label{eq:Mult}
    M(t,\xi)~=~k_{0}(\xi_0/\xi)^\delta(t)^{-\alpha},
\end{equation}
and assume $M~=~10~M_\odot$ and $\Gamma=0.2$. We also compute a reference radius $r_0 = \left[G M (1 - \Gamma)\right]  \hep0^{-1} a^{-2}$ and a reference density $\rho_0 = \mu m_p L_{\text{Edd}} \Gamma \xi_0^{-1} r_0^{-2}$. For the original calculations, the inner radius of the computational domain is $r_{in}~=~r_0$ and the density is $\rho_{in} = \rho_0$. For additional simulations, we explored cases where $r_{in}~>~r_0$ and $\rho_{in}~<~\rho_0$,  see \S-\ref{sec:self}. In addition, 
we assume the adiabatic index $\gamma = 1.0001$, $T = 9.18 \times 10^4$~K, 
and $\xi_{0}~=~{80}$~erg~cm~s$^{-1}$. In these models, we set the multiplier parameters 
$k=0.0076$ and $\alpha=0.742$. We examine wind solutions for various $\delta$ 
values ranging from 0 to 0.4 to investigate the wind solution's dependence on ionization.

We apply the same initial conditions as  \citetalias{Dannen24}.
We also run the simulations in two stages as in \citetalias{Dannen24}: (i) 
we allow the solution to evolve with outflow boundary conditions for 10 to 30 \% of the total runtime and 
(ii) we pause the simulation, modify the outer boundary condition to constant gradients, 
and continue the simulation for the remaining runtime.

\subsection{Data Decomposition and Theoretical Description}\label{sec:M2}

We followed a standard procedure to study the periodic nature of the solutions. 
We decomposed a given wind quantity, $q(t)$, into a time-independent background value, 
$\langle q\rangle$, and time-dependent (Eulerian) perturbation, $q_1(t)$. 
To compute $\langle q \rangle$, we averaged $q$ over approximately a hundred oscillation periods 
using the data from the end of the simulation, when the solution evolved far from the initial transient. 
Then, we subtracted that average from the time-dependent quantities to isolate the oscillatory perturbations, 
i.e., $q(t)~-~\langle q\rangle = q_{1}(t)$.
To simplify notation, we will omit the brackets $\langle q \rangle$ throughout the paper 
and refer to the time-averaged quantities simply as $q$.
To avoid confusion, we write the time-dependent quantity as $q(t)$ when necessary. 

\begin{figure*}[htb!]
    \centering
    \includegraphics[width=0.95\textwidth]{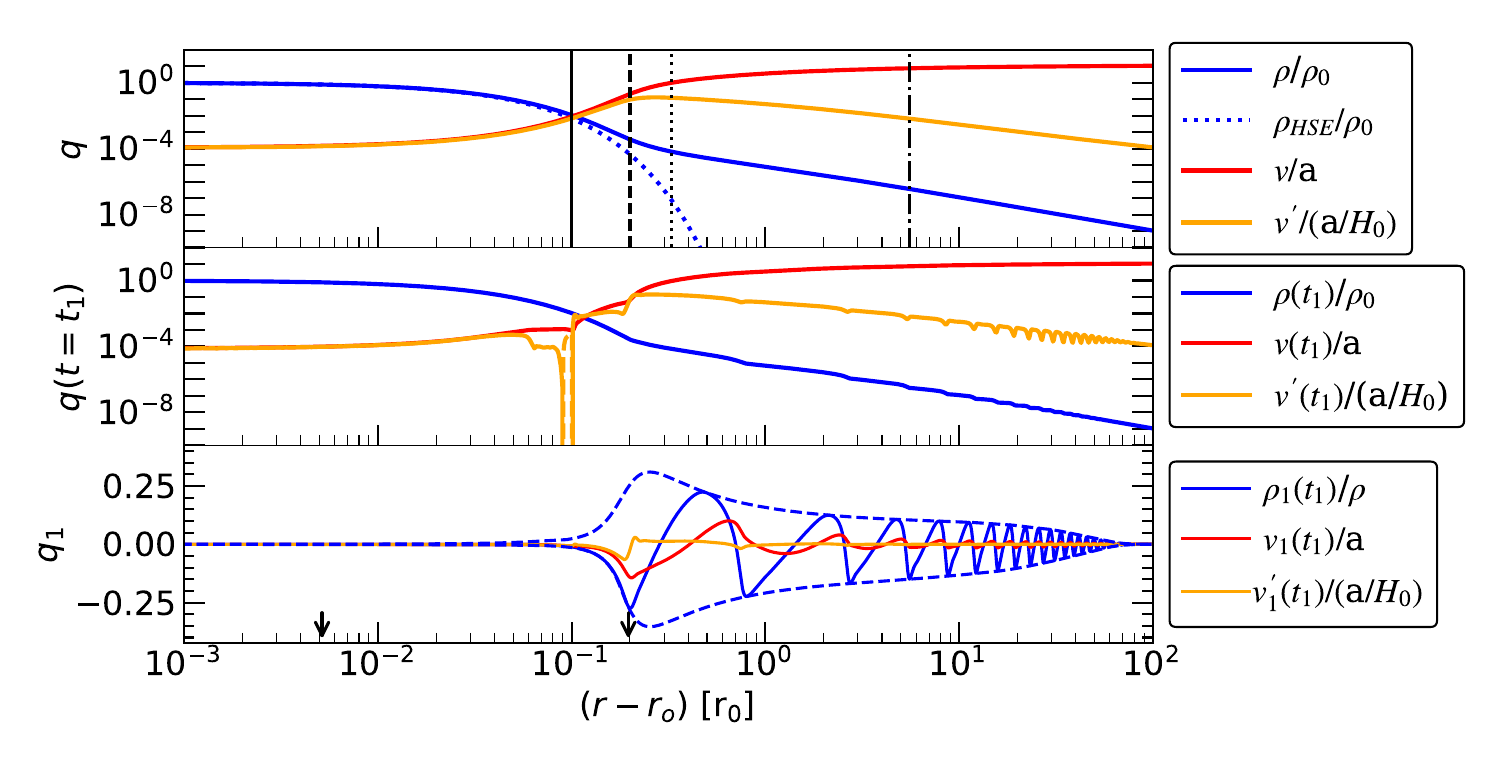}
    \caption{Radial profiles of various wind properties of the $\delta~=~0.3$ periodic solution. 
    We list the specific quantities plotted here in the legends and below.
    The top panel shows the time-averaged profiles, which closely resemble those for the steady-state solutions.
    The black solid and dashed vertical lines mark the outer radius 
    of the unsteady and unstable regions, respectively (see \S~\ref{sec:self} and \ref{sec:w_e} for more detail). 
    The vertical dotted and dash-dotted lines indicate the sonic and critical points, respectively. 
    The middle panel shows an example of time-dependent radial profiles 
    from the simulation at a particular instance $t_1~=~1.59 P_c$. We marked this time in Fig. \ref{fig:Fourier Anlys} 
    and used the data for this to generate the left panels of Fig. \ref{fig:Wave Sol}. 
    The bottom panel shows the perturbed quantities obtained by
    subtracting the data in the middle panel and the time-averaged profiles  
    in the top panel. The two dashed curves represent the envelopes of the perturbed density.
    In the bottom panel, the small vertical arrows at $r~=1.005~r_0$ and $1.196~r_0$ 
    mark the positions we use in the Fourier analysis results presented in Fig. \ref{fig:Fourier Anlys}.}
    \label{fig:Decomp}
\end{figure*}

In the top panel of Fig.~\ref{fig:Decomp}, we show the time-averaged radial profiles 
of density, velocity, and the velocity gradient for the periodic $\delta~=~0.3$ solution. 
These time-averaged profiles do not show any unusual features and 
are similar to the steady-state solutions. 
The middle panel illustrates $\rho(t_1)$, $v(t_1)$, $v'(t_1)$ profiles at a particular time, $t_1$ 
while the bottom panel shows the Eulerian perturbations: $\rho_1(t_1)$, $v_1(t_1)$, and $v'_1(t_1)$. 

\begin{figure*}[ht]
    \centering
    \includegraphics[width=0.95\textwidth]{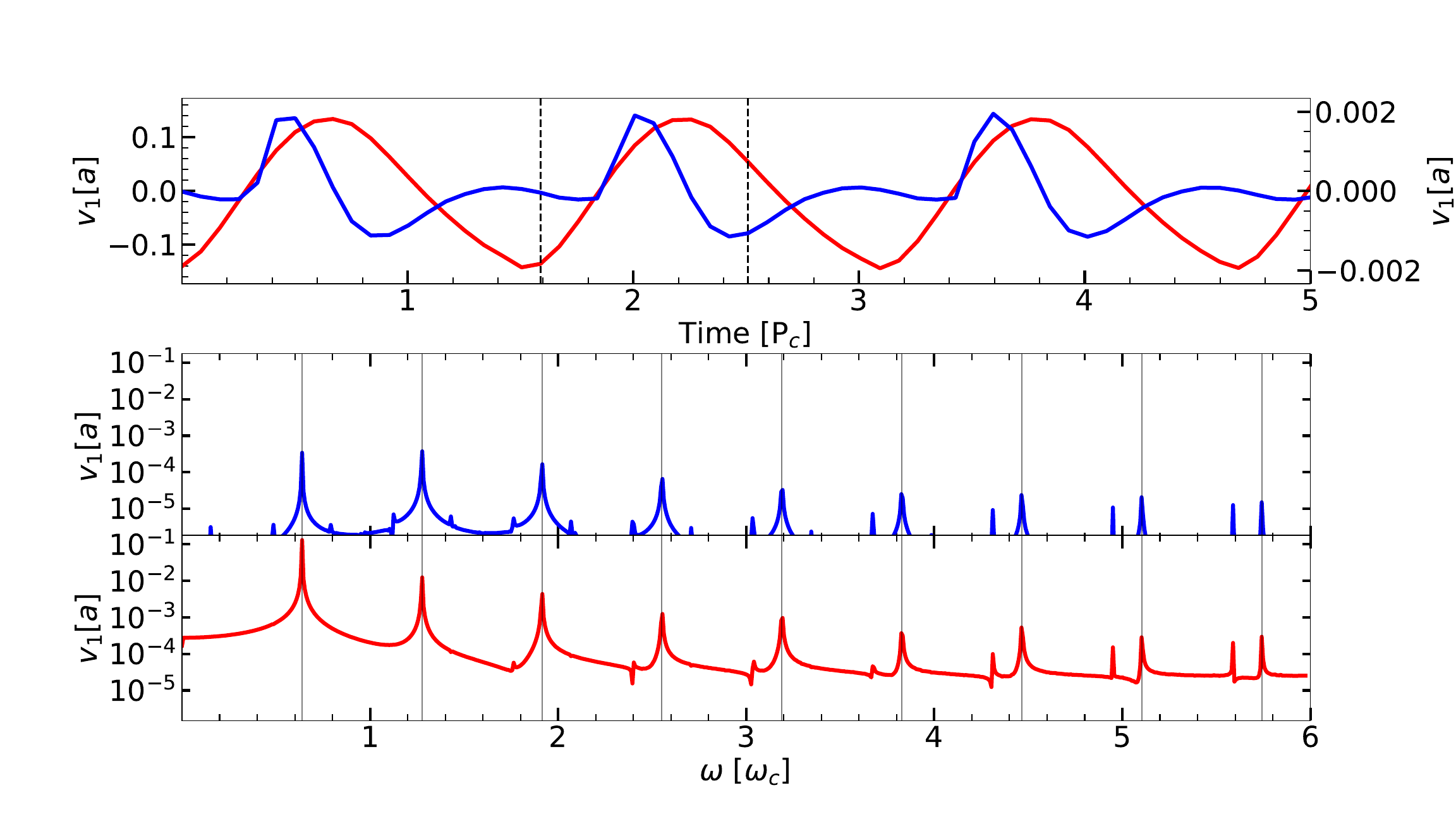}
    \caption{Time variability of the wind properties. The top panel shows examples of the velocity perturbation as a function of time
    at the two radii marked with the small arrows in the bottom panel of Fig. \ref{fig:Decomp} (i.e., at $r~=1.005~r_0$ 
    and $1.196~r_0$, blue and red curves, respectively). 
    The dashed vertical lines in the top panel denote times $t_1$ and $t_2$, where $t_1$ is the time snapshot used in the middle and bottom panels of Fig. \ref{fig:Decomp} and in the left panels in Fig. \ref{fig:Wave Sol}, while $t_2$ is the snapshot used in the right panels of Fig. \ref{fig:Wave Sol}.
    The middle and bottom panel shows the Fourier transform of the temporal signal at  $r~=1.005~r_0$ and $1.196~r_0$, respectively. The vertical gray lines in this pair of panels denote integer multiples of the lowest fundamental frequency, $\omega~=~0.628~\omega_c$. The lines demonstrate that the higher frequencies are higher orders of the fundamental frequency. 
    The middle panel shows that in the unsteady region, the oscillation amplitudes of the three lowest fundamental frequencies are similar to each other and greater than the amplitude of higher-order frequencies.
    The bottom panel shows that the perturbation amplitudes grow rapidly as the perturbations move through the unstable region while maintaining the same frequency structure. Although all frequencies experience growth, the lowest fundamental frequency grows at least an order of magnitude more than the others.} 
    \label{fig:Fourier Anlys}
\end{figure*}

In Fig.~\ref{fig:Fourier Anlys}, we show the results of a Fourier analysis examining 
the temporal properties of the perturbations. In the top panel, we present the temporal signal 
at $r~=~1.005~r_0$ and $r~=~1.196~r_0$, blue and red curves, respectively. We indicated 
these two locations with downward-pointing arrows in the bottom panel of Fig.~\ref{fig:Decomp}. 
In the bottom panel of Fig.~\ref{fig:Fourier Anlys}, we present the frequency power spectrum, 
which shows a clear preferential amplification of the fundamental mode at $\omega~=~0.628~\omega_c$. 
We will revisit these two figures in the next section. 
  \hspace{0.02\textwidth} 
To quantify the wave behavior seen in the periodic solutions, we measure the wave energy and its evolution
and examine the solutions of a mass-flux wave equation.
In Appendix~\ref{sec:A1},
we derive the wave energy conservation equation 
\begin{multline}
  \label{eq:CAE}
  \frac{\partial E_w}{\partial t} + \frac{1}{r^{2}} \frac{\partial(r^{2} p_1v_{1})}{\partial r}~=~\\ - \rho v_{1}\frac{\partial( v v_{1})}{\partial r}~-~\frac{\rho_{1}}{ \rho } \frac{1}{r^{2}} \frac{\partial(r^{2} p_1 v)}{\partial r} +  \rho v_{1}g_{l,1},
\end{multline}
where $E_w$ is  the specific acoustic energy  defined 
as the sum of the potential energy component, $PE_w$ and the kinetic energy component, $KE_w$,
\begin{equation} 
  \label{eq:AE}
    E_w~\equiv~PE_w~+~KE_w~=~\frac{1}{2}\frac{p_1^{2}}{ \rho a^{2}} + \frac{1}{2} \rho v_{1}^{2}.
\end{equation}
The perturbed line force, $g_{l,1}$, can be expressed as
\begin{equation}
  \label{eq:frad1Va}
    g_{l,1}~=~2v_{a}\left[\left(\frac{\delta}{\alpha} -1\right)\frac{\rho_{1}}{\rho}  v'  + v_1'\right],
\end{equation}
where $v_a$ is the Abbott speed that depends 
on the ratio between the background line force and the velocity gradient:
\begin{equation}
  \label{eq:Va}
    v_a~\equiv~\frac{1}{2} g_l' =\frac{1}{2}\frac{\partial  g_l}{\partial \left(  v' \right)}~=~\frac{\alpha}{2} \frac{ g_l}{ v'}.
\end{equation}
The right-hand side (RHS) of eq. \ref{eq:CAE} contains the sink 
and source terms that either add or remove energy from the acoustic wave. In a static, homogeneous fluid, these terms equal zero and the flux of energy density entering 
or leaving a location $r$ drives the time rate of change of $E_w$.

In Appendix~\ref{sec:A2}, we derived the mass flux wave equation
for an isothermal mCAK line-driven wind:
\begin{multline}
  \label{eq:Wave}
    \frac{\partial^2 f_1}{\partial t^2} + 2v_{\text{eff}}\frac{\partial^2 f_1}{\partial r\partial t}  + \left( v_{\text{eff}}^2~-~a_{\text{eff}}^2\right)\frac{\partial^2 f_1}{\partial r^2}~=~\\~+~\frac{~a_{\text{eff}}^2-3v_{\text{eff}}^2~}{\lambda_v}\frac{\partial f_1}{\partial r}~+~\frac{2v_a\left( v_a~-~v\left(\frac{\delta}{\alpha}\right) \right)}{\lambda_v}\frac{\partial f_1}{\partial r} \\~-~\frac{2v_{\text{eff}}}{\lambda_v}\frac{\partial f_1}{\partial t}, 
\end{multline}
where $v_{\text{eff}}~=~ v~-~v_a$, $a_{\text{eff}}^2~=~a^2 + v_a^2$, and $\lambda_v~=~v/v'$, and the mass flux perturbation defined as 
\begin{equation}
    \label{eq:pertMCf}
    f_{1}~\equiv~  f \left(\frac{\rho_{1}}{ \rho} + \frac{v_{1}}{ v}\right).
\end{equation}
The mass flux $f~=~r^2\rho v$ is constant.
Eq. \ref{eq:Wave} reduces to the Abbott wave equation when one neglects gravity,  
stratification, and divergence
\citep[see eq. 42 in][]{abbott1980theory}. We also note that when $v_a=0$,
eq. \ref{eq:Wave} reduces to the isothermal acoustic wave equation
\cite[see eq. 6 in][]{1980MNRAS.191..571P}. 

\section{Results}\label{sec:results}

As mentioned in \S~\ref{sec:intro}, we identified an instability that amplifies radiative-acoustic waves
in a small region inside a strongly stratified, nearly hydrostatic atmosphere.
We found that the subsonic highly stratified part of each periodic solution 
has two distinct regions: 
(i) unsteady region where perturbations are self-excited 
and (ii) unstable region where perturbations are amplified
In the top panel of Fig. \ref{fig:Decomp}, we use the vertical solid and dashed lines to mark
the outer radius of these two regions (we also mark the sonic and critical points using
the dotted and dash-dotted lines).

To ascertain the source of the instability, we begin in \S~\ref{sec: CtPt AS} by discussing 
how the time-averaged flow properties depend on $\delta$. In \S~\ref{sec:w_e} we discuss 
the flow properties responsible for the instability and how they lead to 
a wave amplitude that grows beyond that predicted by conserving wave energy flux 
in a stratified atmosphere.  
In \S~\ref{sec:Instability} and \S~\ref{sec:self}, we return to our discussion of the amplification and source of the perturbation. Unless stated otherwise, we describe the behavior of the solution for $\delta = 0.3$. 
The qualitative features of the time-dependent behavior observed in the $\delta = 0.3$ solution are similar to those found across periodic solutions within the range of $\delta$ from 0.23 to 0.4.
\subsection{mCAK Critical Point and Abbott Speed}\label{sec: CtPt AS}
The emergence of periodic oscillations in the wind solutions for \(\delta > 0\) are related to changes in the characteristics of the background flow, specifically the magnitude of the velocity and velocity gradient near the inner boundary. Qualitatively this makes sense in the context of the analogy from \cite{owocki2020getting}. In the line driven winds of OB stars, the line force drives material away from the star creating a suction that allows the material from the hydrostatic region to expand upward through the sonic point. By increasing the ionization dependence via $\delta$ and weakening the line force, the suction effect lessens, leaving the material in an increasingly hydrostatic state (reduced velocity and velocity gradient).  

In early work, researchers modeled line force weakening with $\delta$ parameters ranging from 0-0.16 and found agreement with predicted and observed mass loss rates of OB stars \citep{LC}. More recently, \cite{Curé_2011} studied line driven winds for a large range of $\delta$ ($0-0.42$) and found the emergence of a new type of stationary solution, the $\delta-$slow solution. The $\delta-$slow solutions exhibit slower terminal velocities and mass loss rates around a 100 times lower than the fast wind solutions, in agreement with observations of Type-A supergiants. Additionally, \cite{Curé_2011} found no slow wind solutions in the range $\delta = 0.23-0.30$. Larger values of $\delta$ are typically unrealistic for OB stars, however for hydrogen winds \cite{Puls_2000} showed that $\delta\sim0.3$, and \cite{kudritzki2002line} demonstrated that low metallicities significantly increases the number of ionizing photons in the wind. In the context of AGN and XRBs, \cite{Dannen24} calculated $k, \alpha$, and $\delta$ and found that $\delta$ can approach values of $\alpha$ in line driven winds.

\begin{figure*}[ht]
  \centering
  \includegraphics[width=0.85\textwidth]{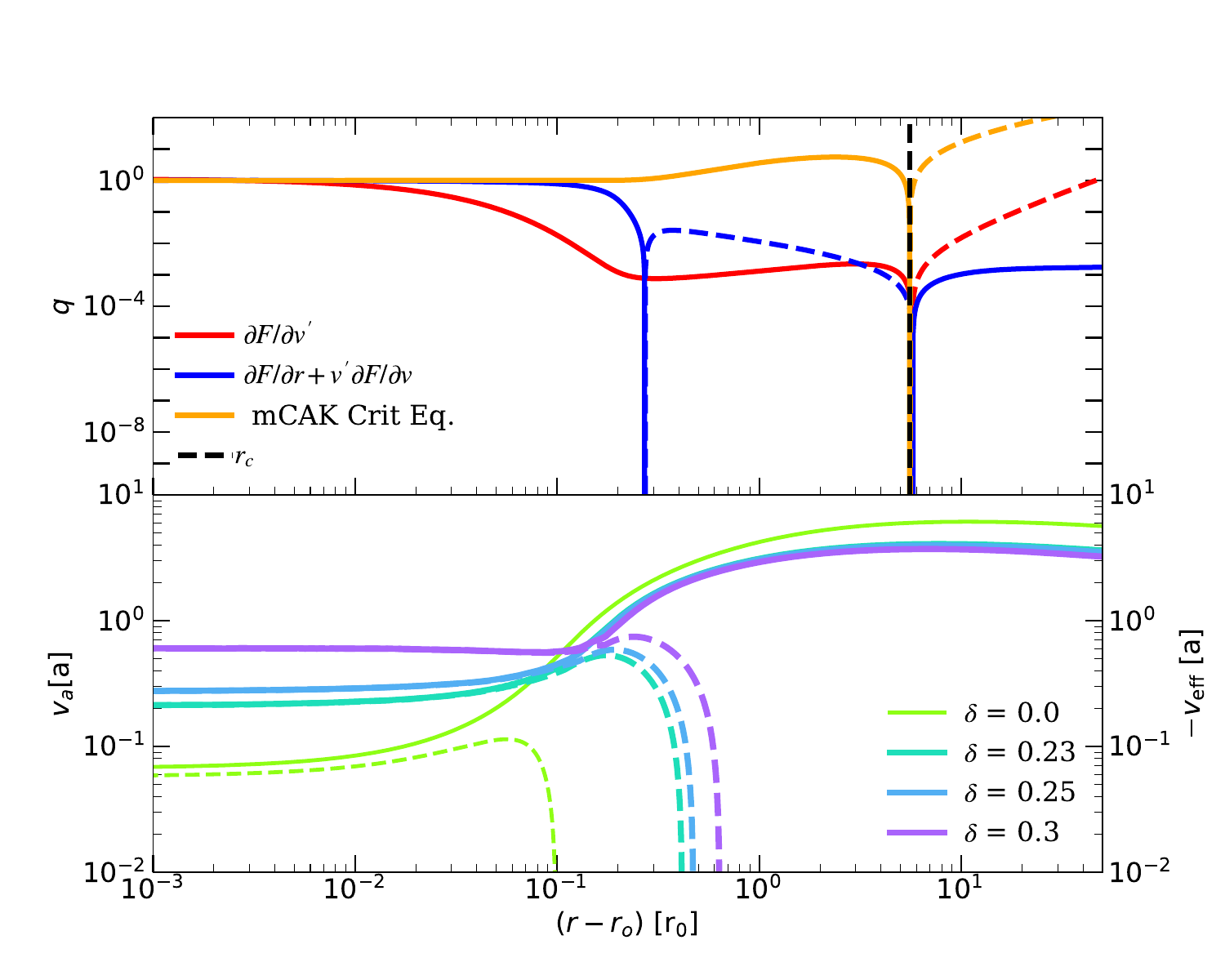}
  \caption{Comparison of radial profiles of wind properties.
 The top panel shows the location of the critical point, $r_c$, for the $\delta~=~0.3$ solution. 
 The vertical dashed black line marks the location where
 the LHS of eq.~\ref{eq:mCPt} (orange curve) equals zero. We confirm that at $r_c$ as determined 
 by eq~\ref{eq:mCPt} coincides with the radius where the regularity and singularity conditions are satisfied, i.e.,  
 where both eq.~\ref{eq:regeq}  (red) and eq. \ref{eq:singeq} (blue) equals zero.
 All curves in the top plot are normalized to equal one at $r~=~r_0$. 
 The bottom panel illustrates how the weakening of the line force (i.e., an increase in $\delta$) changes the wind properties.
 The solid curves correspond to the Abbott speed, $v_a$ 
 (refer to the ordinate axis of the left and see eq. \ref{eq:Va} for the definition of $v_a$) while the dashed curves correspond
 to the negative value of the effective velocity, $v_{\text{eff}}=v-v_a$ (refer to the ordinate axis of the right).
 The effect of increasing $\delta$ is the strongest at the smallest radii, where the stratification is the highest. 
 The increase/decrease of the Abbott speed/flow velocity with $\delta$ results in the increase of the size of the inner region where $v_{\text{eff}}<0$. This size is five times larger, while $v_a$ is nearly an order of magnitude greater for $\delta=0.3$ compared to $\delta=0$. Note that for small radii and high $\delta$, the solid and dashed curves overlap because $v/v_a<<1$ and $-v_{\text{eff}} \simeq v_a$.
}  
\label{fig:CritPt}
\end{figure*}

To investigate evolution of background properties specific to our line driving parameters 
$\xi_{0}~=~{80}$~erg~cm~s$^{-1}$, $k=0.0076$ and $\alpha=0.742$, we follow \citetalias{CAK}'s 
mathematical analysis of the steady-state momentum equation expressed as 
\begin{multline}
  \label{eq:momeqND}
    F(r,v,v')~\equiv\\~1 + w' +\frac{1}{2\hep0}\frac{w'}{w} + \frac{2}{\hep0}\frac{r}{r_0}~-~Ch(|w'|)^\alpha~=~0,  
\end{multline}
where $w~=~v^2/[(1-\Gamma)v_{esc}^2]$, $w'=\frac{\partial w}{\partial n}$, $n~=~1-r_0/r$,  
and $h$ is a dimensionless correction factor that could related to the finite disk correction 
or ionization effects. In our models, this factor is purely due to the latter, 
hence  $h=(\xi_0/\xi)^\delta$. Using the function $F(r,v,v')$ 
and time-averaged quantities, we compute the regularity equation, 
\begin{equation}
  \label{eq:regeq}
    F_r \equiv \frac{\partial F}{\partial v'},
\end{equation}
and the singularity equation, 
\begin{equation}
  \label{eq:singeq}
    F_s \equiv \frac{\partial F}{\partial r} + v'\frac{\partial F}{\partial v}.
\end{equation}
We plot these two functions using red and blue curves in the top panel of Fig.~\ref{fig:CritPt}. 
The critical point of the wind solution is where both functions are equal to zero, 
and it occurs at $r_c~=~6.55~r_{0}$ (see the vertical black dashed line 
in the top panel of Fig.~\ref{fig:CritPt};
we also marked  this point with the vertical dashed-dotted line in the top panel 
of Fig.~\ref{fig:Decomp}). 
Additionally, we derived an explicit relation for how $r_c$ depends on $\delta$:
\begin{equation}
  \label{eq:mCPt}
    1~-\frac{\delta}{2\alpha}~-\left( \frac{v_c}{r_c v_c'} \right)^2 \left( 1~-~\frac{\delta}{2\alpha} \left( \frac{r_c v_c'}{a} \right)^2 \right)~=~0.
\end{equation}
This equation reduces to the classical CAK critical point condition, $v_c/r_c= v'_c$,  
for $\delta=0$. In our subsequent literature search, we found work by \cite{cure2023radiation} 
and that their critical function  (eq.~57) simplifies to our eq. \ref{eq:mCPt} when one
neglects the effects of the finite disk correction factor and rotation.  

The $v_c/r_c= v’_c$ condition only implies a slight outward shift in the critical point 
due to the weakening of the line force via $\delta$. By including the critical point's 
explicit dependence on $\delta$, i.e., using eq. \ref{eq:mCPt}, we show that the critical 
point shifts significantly. In the $\delta~=~0$ solution, the critical point 
is at $r_{c}~=~1.93~r_{0}$, while for the $\delta~=~0.3$ solution, the critical point 
is farther out by a factor more than 3, i.e., $r_c~=~6.55~r_{0}$. 
We verified eq.~\ref{eq:mCPt} with calculations of eq.~\ref{eq:regeq} and eq.~\ref{eq:singeq} 
(see the top panel of Fig. \ref{fig:CritPt}).

Unlike $r_c$, we found that $v_c$ does not strongly depend on $\delta$.
Consequently, the wind acceleration decreases with increasing $\delta$. 
This trend also influences another characteristic of line-driven winds: the Abbott speed
(see the bottom panel of Fig. \ref{fig:CritPt}.) We can describe 
the increase of $v_a$ with $\delta$ by  the following scaling:
\begin{equation}
  \label{eq:Va-prop}
    v_a~\propto \frac{v^{-\delta} v^{\prime\alpha}}{v' \rho^\alpha} \propto \frac{v^{\prime \alpha - \delta -1}}{\rho^\alpha}
\end{equation}

The density dependence, $\rho^{-\alpha}$, arises from the line-force dependence of the density 
through the optical depth. In HSE, gas density does not appreciably depend on $\delta$. Therefore, 
the change in $v_a$ induced by modifying $\delta$ comes only through $v$ and $v'$.
The velocity dependence, $v^{-\delta}$, stems from the ionization dependence 
and the velocity gradient dependence, $v^{\prime\alpha}$, comes from the optical depth 
in eq. \ref{eq:Frad}. 
At small radii, as $\rho$ decreases exponentially, $v$ increases exponentially, and
$v \propto v'$ (compare the red and orange curves in the top and middle panels
of Fig.~\ref{fig:Decomp}),
leading the final expression
on the RHS of eq. \ref{eq:Va-prop}.
The net result is almost an order of magnitude increase in $v_a$ 
for nearly two orders of magnitude decrease in $v$ and $v'$.
Thus, for the $\delta = 0.3$ solution, the stratified region increasingly becomes  
hydrostatic whereas $v_{\text{eff}}$ becomes an order of magnitude more negative 
across a region five times larger than that of the $\delta = 0$ solution
(see the dashed curves in the bottom panel of Fig. ~\ref{fig:CritPt}). 
Generally, as $\delta$ increases, the outer radius of the negative $v_{\text{eff}}$ 
region increases. In a purely Parker wind, $v_{\text{eff}}$ would always be positive, 
so we attribute the extension of this negative region to the growth of radiative acoustic waves. 
We provide more support for this below.

\begin{figure*}[ht]
  \centering
  \includegraphics[width=1\textwidth]{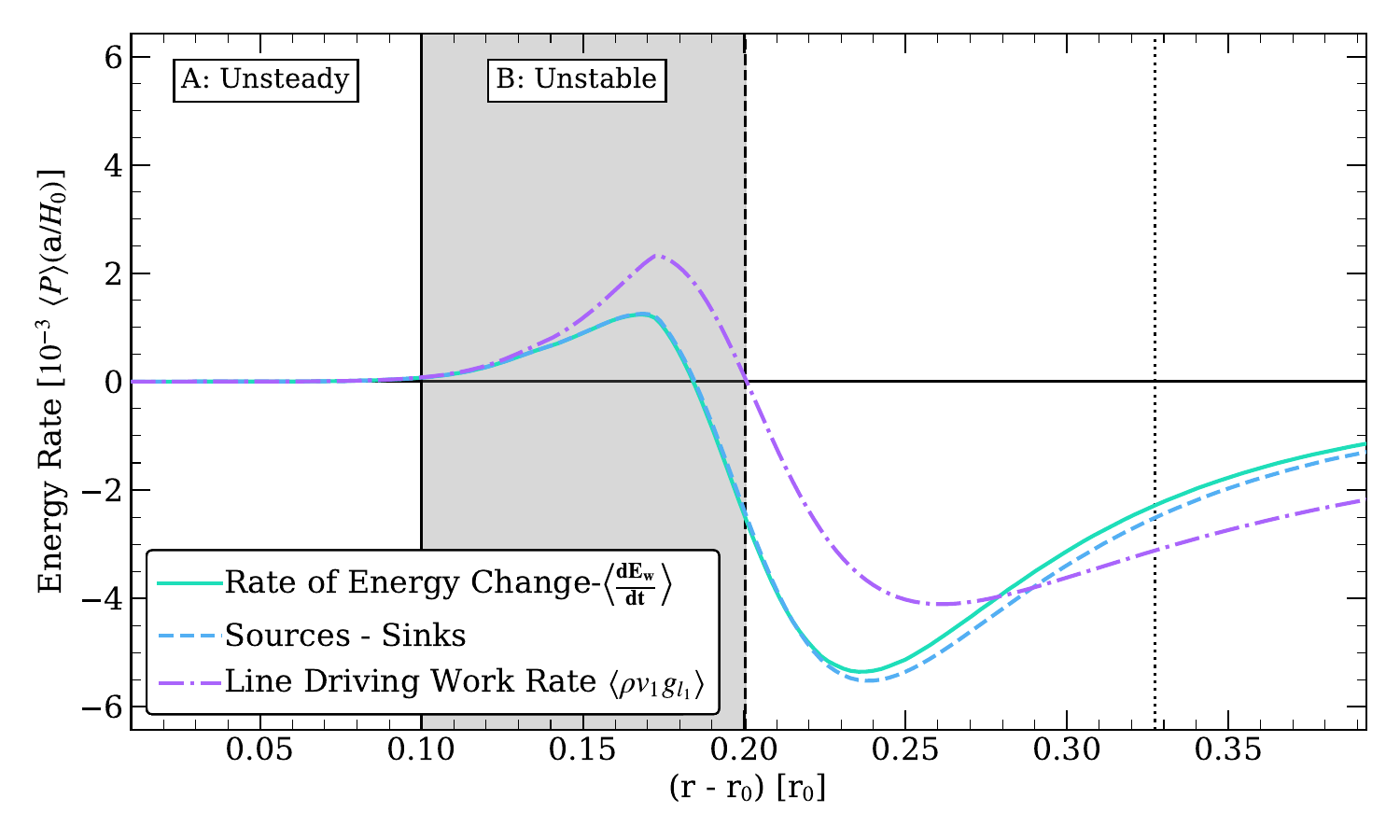}
  \caption{Radial profiles of the time-averaged terms in the wave energy equation, eq. \ref{eq:CAE} 
  (see the legend in the bottom left corner). The dashed vertical line marks the radius 
  where the line force's work rate equals zero. We use this radius to delineate the boundary between 
  the highly stratified atmosphere and the launched wind. We refer to the former and latter 
  as unstable and transonic regions. The solid black line, determined empirically (see~\S~\ref{sec:self}), 
  marks the boundary between the intrinsically time-dependent flow and the unstable region (shaded in gray). We refer to the intrinsically time-dependent region as the ‘unsteady’ region. In this region, self-excitation generates waves that grow 
  in the unstable region. The dotted line marks the sonic point. 
  Note that all relevant activity is in the subsonic, almost static, region.
}
  \label{fig:SWPower}
\end{figure*}

\subsection{Wave Energy} \label{sec:w_e}

We use eq.~\ref{eq:CAE} and simulation data to compute the time-averaged wave energy rate 
and the corresponding source/sink terms. In Fig.~\ref{fig:SWPower}, the solid curve shows 
the average wave energy rate, $\langle\frac{d E_w}{d t}\rangle$, i.e.,  the LHS of eq. eq.~\ref{eq:CAE}. 
In the unsteady region, the rate is positive but relatively small at the smallest radii. 
However, the rate dramatically increases farther downstream in the unstable region and 
reaches a maximum at $r~\simeq~1.17~r_0$.  This increase directly corresponds to the sharp increase 
in the perturbation amplitudes, as shown in the bottom panel of Fig.~\ref{fig:Decomp}. 
Even farther downstream, the rate decreases and becomes negative at $r~\simeq~1.85~r_0$.
The dashed curve represents the source and sink terms on the RHS of eq. \ref{eq:CAE}. 
Both curves agree, confirming the expectation from the linear analysis that the source/sink 
terms directly drive the changes in the wave energy rate of change. 
The dash-dotted curve shows only the line driving term on the RHS of eq. \ref{eq:CAE}, $\langle\rho v_1 g_{l,1}\rangle$. 
We found that this term is responsible for depositing energy into the wave at small radii. It is positive
up to the radius of $1.2~r_0$, which we defined as the outer radius of the unstable region.

Based on scaling relations of a linear pressure wave, we wouldn't expect a positive contribution from  $\langle\rho v_1 g_{l,1}\rangle$.
We can express the power produced by the perturbed line force in terms of $\rho_1$ and $v_1$ as
\begin{equation}
    \label{eq:line force 1}
    \rho v_{1} g_{l,1} = ~2v_{a} \left[\left(\frac{\delta}{\alpha} -1\right)\rho_{1}v_{1} v'  + \rho v_{1}v_{1}'\right].
\end{equation}
For a linear acoustic perturbation $\rho_1 \propto v_1$ 
whereas $i(2\pi/\lambda)v_1 \propto v’_1$. 
Therefore, the $\rho_1 v_1$ contribution would indicate energy removal
and $v_1 v'_1$ would contribute no net effect. 
Thus, the net contribution of all three terms on the RHS of eq.~\ref{eq:CAE} should be negative
when considering a local oscillatory perturbation (for more detail, 
see Appendix~\ref{sec:A3} and Key et al., submitted). 

The clue to the positive value of $\langle\rho v_1 g_{l,1}\rangle$ comes from 
Fig.~\ref{fig:Decomp} which shows $v_1$ is nearly proportional to $v'_1$ 
in the stratified region. 
In this region, the WKB approximation does not apply
because the spatial component of the phase, $\phi$, changes over a similar 
scale as the perturbation amplitude, $\abs{q_1}$, i.e.,  
\begin{equation}
  \label{eq:WKB }
   \abs{\frac{d \phi(r)}{dr}} \approx \frac{\abs{q'_1}}{\abs{q_1}}. 
\end{equation}
Obtaining analytical results in this regime is challenging and beyond the scope 
of this work. 

In the simplified case of 
an acoustic wave in a hydrostatic stratified atmosphere, one can show that
$v_1 \propto \exp(z/2H)\exp[i(2\pi/\lambda)z - \omega t)]$ and $\lambda~=$ constant, 
therefore  $\langle v_1 v'_1 \rangle \neq 0$. 
We note that in our simulations, the line force ceases to perform positive work 
at the end of the stratified region. This transition is ubiquitous across the periodic simulations with
$0.23 \leq \delta \leq 0.4$, indicating the exponentially stratified atmosphere 
is critical for wave energy growth. We elaborate on this point by considering the wave equation, eq. \ref{eq:Wave}.

\subsection{Instability} \label{sec:Instability}

All terms on the RHS of  eq.~\ref{eq:Wave} depend on the fluid stratification 
as they contain $\lambda_v$ in the denominator. 
Thus, $\lambda_v$ sets the frequencies characterizing the stratification and line driving: 
(i) the acoustic cut-off frequency, $\omega_c = a/2\lambda_v$ and (ii) the line-driving frequency $\omega_{\text{eff}} \equiv -v_{\text{eff}}/\lambda_v \simeq v_a/\lambda_v$. Because $v$ is nearly zero ($v/a < 10^{-1}$) and $v_{\text{eff}}$, $v_a$ and  $a_{\text{eff}}$ are nearly constant in much of the unstable region, we can simplify eq. \ref{eq:Wave} to 

\begin{multline}
  \label{eq:WaveHS}
    \frac{\partial^2 f_1}{\partial t^2} - 2v_{a}\frac{\partial^2 f_1}{\partial r\partial t}  -  a^2\frac{\partial^2 f_1}{\partial r^2}~=~\\~2 \omega_c a\frac{\partial f_1}{\partial r}+~2 \omega_{\text{eff}}\frac{\partial f_1}{\partial t},
\end{multline}
In our simulation, $v_{a}~\simeq~0.6~a$, so  $a_{\text{eff}}~\simeq~1.2~a$ and $\omega_{\text{eff}}~\simeq~1.2~\omega_c$. 
To compare the characteristic length scale of the line driving, we recast eq. \ref{eq:WaveHS} using the normal variables $x~=~r - (v_{\text{eff}} + a_{\text{eff}})t$ 
and $y~=~r + (-v_{\text{eff}}~+~a_{\text{eff}})t$:
\begin{equation}
  \label{eq:WavePQ}
  -4\frac{\partial^2 f_{1}}{\partial x\partial y} = \frac{A_1 + A_2}{\lambda_v} \frac{\partial f_1}{\partial x} + \frac{A_1 - A_2}{\lambda_v} \frac{\partial f_1}{\partial y},
\end{equation}
where 
\begin{equation}
  \label{eq:A1}
   A_{1} \approx 1~+~\frac{v_a^2}{a_{\text{eff}}^2},  
\end{equation}
and
\begin{equation}
  \label{eq:A2}
   A_{2} \approx 2\frac{v_{a}}{a_{\text{eff}}}. 
\end{equation}
For $\delta = 0.3$, $(A_1 + A_2)~\simeq~9/4$ and $(A_1-A_2)~\simeq~1/4$. 
Without line driving, $(A_1 + A_2) = 1$ and $(A_1 - A_2) = 1$. Given the factors $A_1$ and $A_2$ are comparable, 
we expect the line driving and stratification effects to operate on a similar length scale. 
We note that the line force not only breaks the symmetry in the propagation speeds of the acoustic modes
but also breaks the symmetry between the characteristic lengths for the inward and outward modes in a stratified atmosphere.

Based on the similar characteristic lengths and frequencies between the stratified acoustic behavior 
and the line force, we expect qualitatively different behavior for frequencies above $\omega_c$ and below $\omega_c$, which we see in Fig. \ref{fig:Fourier Anlys}.
We note that $\omega_{\text{eff}}$ will approach zero and become negative 
as $v$ approaches and exceeds $v_a$, occurring outside the exponentially stratified region near $r~\simeq~1.6~r_0$. Crucially, $\omega_c$ and $\omega_{\text{eff}}$  will both approach zero as $\lambda_v$ increases, reinforcing the conclusion of \S~\ref{sec:w_e} that the exponentially stratified atmosphere is critical to wave energy growth for the periodic simulations.

To gain further insight into how line driving increases $\rho_1$ and $v_1$, we return to the general case of  \ref{eq:Wave}. We numerically solve eq. \ref{eq:Wave} for the dominant frequency observed in the simulations.

For completeness, we briefly describe our method below.
We substitute $f_1~=~g(r)h(t)$ where $h(t)~=~\exp[-i\omega t]$ into eq. \ref{eq:Wave}, 
decomposing the wave equation into the complex second-order ODE:
\begin{multline}
  \label{eq:ODE}
   \left( \omega^2 + \frac{2i \omega v_{\text{eff}}}{\lambda_v} \right) g(r)
    \\ + \left( \frac{~a_{\text{eff}}^2~+~2 v_a \left( v_a~-~v \frac{\delta}{\alpha} \right)~-3 v_{\text{eff}}^2}{\lambda_v}~+~2i \omega v_{\text{eff}} \right) g'(r)
    \\ + \left( a_{\text{eff}}^2~-~v_{\text{eff}}^2 \right) g''(r)~=~0.
\end{multline}
We force a continuous, purely real oscillation frequency at the inner radius using the above substitution. 
Physically, the continuous oscillation would be analogous to having a steady stream of acoustic waves entering the domain.

The imposed real frequency forces 
amplitude changes to depend on the imaginary components of the spatial solution, $g(r)$. 
We solved eq. \ref{eq:ODE} by splitting the 2nd-order ODE into a system of first-order ODEs, 
using a Runge-Kutta solver. 
We applied the Neumann and Dirichlet boundary condition for an outward propagating homogenous eigenmode, and updated 
the coefficients of the ODE at each grid point using the background quantities supplied 
by the time-average flow properties. We apply a perturbation amplitude of $~2.7\times 10^{-3}$ 
for both $\rho_{1}/\rho$, and $v_{1}/a$ at the wind base, and set the frequency to $\omega~=~0.628~\omega_c$. 
With the solution of $f_1$, we extract $\rho_1$ and $v_1$ by using eq. \ref{eq:PertMC} 
and eq. \ref{eq:pertMCf} to obtain the complete acoustic wave behavior of the system.

\begin{figure*}[ht!]
  \centering
  \includegraphics[width=1\textwidth]{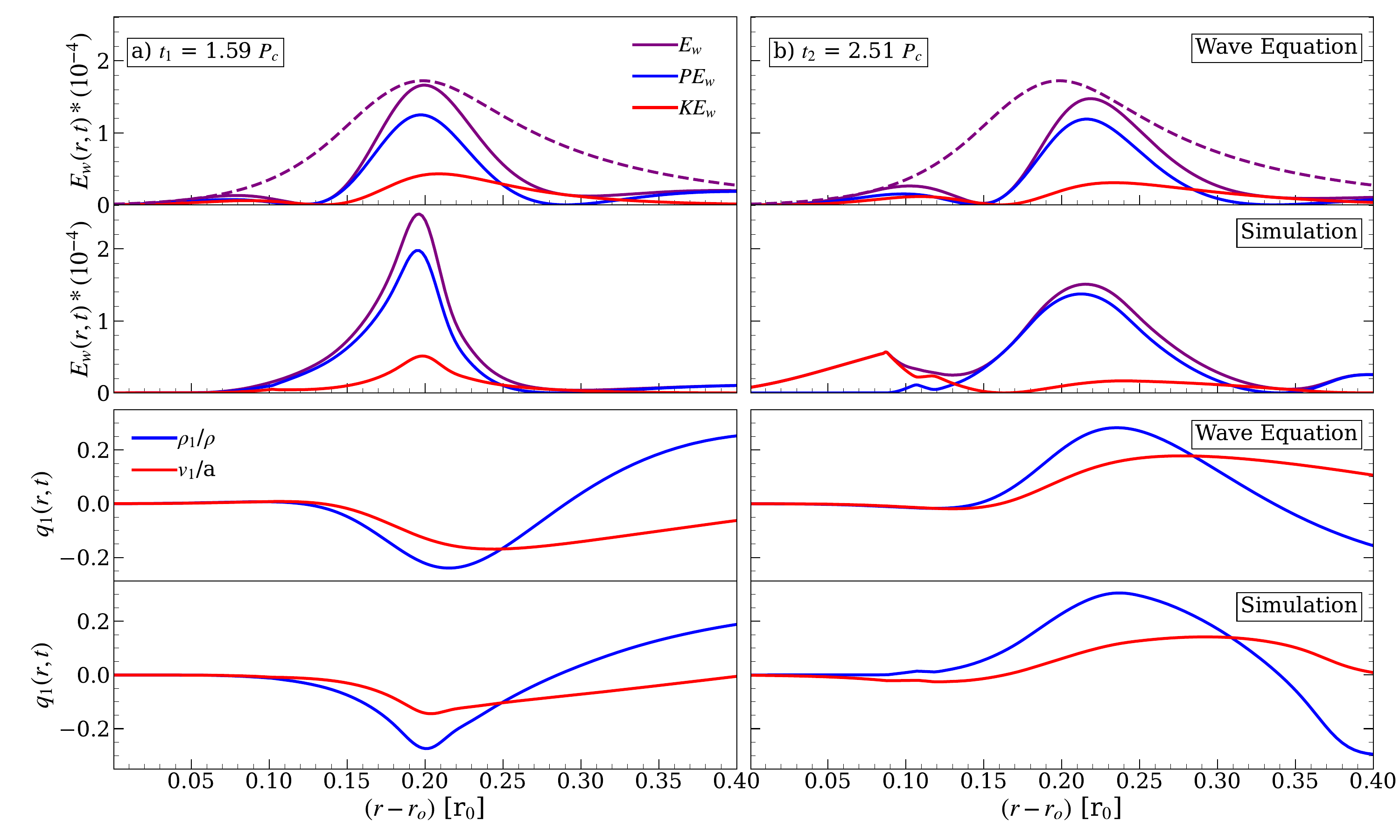}
  \caption{Comparison between the solution of the wave equation for a single-dominant frequency 
and the multi-frequency wave obtained from the time-dependent simulation
(see the legends in the top left corner of the panels in the right column). The top two panels show radial profiles of $E_w$ and its potential ($PE_w$) and kinetic ($KE_w$) components,
the purple, blue, and red curves, respectively (see eq.~\ref{eq:AE}).  
The multi-frequency behavior of the time-dependent simulation is illustrated in
Fig. \ref{fig:Fourier Anlys}.
The bottom two panels show corresponding density and velocity perturbation profiles,  
blue and red curves, respectively. The panels in the left column correspond to the time $t_1~=~1.59 P_c$
where the wind starts its periodic cycle. The panels in the right column show a specific snapshot where the solution experiences a kink after a time $\Delta t~=~0.92~P_{c}$.
Both $t_1$ and $t_2$ are marked with the vertical dashed lines in the top panel of Fig.~\ref{fig:Fourier Anlys}.
  }
  \label{fig:Wave Sol}
\end{figure*}

We present the results of our calculations in Fig.~\ref{fig:Wave Sol}. 
In the left column, we show a time snapshot of the simulation results used in Fig. \ref{fig:Decomp} 
and the corresponding time snapshot from the linearized solution. 
In this snapshot, the linearized solution exhibits similar amplitude and phase behavior 
for $\rho_1/\rho$ and $v_1/a$ compared to the simulation results. 
Additionally, the linearized solution shows similar energy growth and distribution 
among kinetic and potential wave energy. 
In the right column, we show that in the time-dependent simulation, there is a time during
the periodic cycle when eq.~\ref{eq:Wave} does not adequately capture the periodic behavior.
Specifically, we find
a snapshot containing a kink in the unsteady flow region in the simulation results. 
The kink is a discontinuous change in the slope of $v_1$ corresponding to a discontinuous jump in $v_1'$. 
We determined that the kink is caused by the absolute value of $v'$ in the force multiplier prescription. 
We will discuss the kink and its significance in \S~\ref{sec:self}. 

Overall, we found that oscillatory behavior originates in the unsteady region. 
The solution continuously oscillates in this region, injecting fluctuations 
into the rest of the domain. As fluctuations propagate, they evolve in shape and amplitude, 
becoming Abbott waves that can be, to some degree, captured by eq. \ref{eq:Wave}. 

Three factors limit the spatial and temporal similarity between the linearized solution and the simulation. 
First, the wave in the simulation contains multiple frequencies (shown in Fig. \ref{fig:Fourier Anlys}). 
Second, the simulation results include the second-order effects neglected in the linearized solution. 
Lastly, the linearized solution is not a pure outward propagating wave. At frequencies comparable 
to $\omega_c$, there is coupling between the homogeneous outward and inward propagating eigenmodes 
of acoustic waves in an isothermal wind \citep{grappin1997acoustic}. Therefore, using the homogeneous 
outward propagating boundary conditions does not yield a pure outward propagating wave.

In the linearized solution, we found that the energy grows by a factor of $\sim$ 256 
from the inner boundary to the end of the unstable region. This growth factor indicates 
an average growth scale length  $\sim 1.8\times \lambda_v$. We also found that this behavior directly 
results from the dependence of the line force on $v'$. 

We performed wave solution calculations for different model parameters, including a case where we set $\delta = \alpha$.
In this case, the $\rho_1$ contribution to the line force in the linearized solution equals zero (see eq. \ref{eq:frad1Va}). 
The remaining terms in the wave behavior are due to the perturbed pressure and 
and velocity gradient in the perturbed line force.  We found that the $\delta = \alpha$ solution
is qualitatively similar to the solution in Fig. \ref{fig:Wave Sol} but has a slightly higher energy maximum 
and a shallower energy drop-off as the wave propagates outwards. 
This difference in behavior indicates that the $\rho_1$ term removes energy from the wave via the line force.
It is also consistent with the expected behavior discussed in \S~\ref{sec:w_e}: 
the $\langle\rho_1 v_1\rangle$ term leads to the energy removal.

We conclude that the positive energy contribution from the line force must be 
from the term $\langle v_1 v'_1 \rangle  > 0$ and a non-WKB behavior.
In an exponentially stratified atmosphere, velocity perturbation amplitudes increase by conserving wave energy flux \citep{clarke2007principles}. 
The increase induces steeper gradients, amplifying the line force and, subsequently, the wave. The feedback between the two effects continues until the specific energy sink rate from spherical divergence, 
wave-flow interaction, and the density contribution to the line-driving perturbation 
becomes significant.

\subsection{Self-Excitation}\label{sec:self}

As mentioned in \S~\ref{sec:Instability}, the periodic time-dependent simulations show the reappearance of a distinctive kink in $v_1$ 
in the innermost part of the flow that we call the unsteady region ($r~<~1.1~r_0$ and $v~<~0.01~a$). The kink emerges when the line force cannot steadily accelerate the outflow. 
This situation can arise when the mass flux of the wind is larger than the line force can support. 
Thus, the outflow must slow down and potentially turn into an inflow \cite[e.g.,][]{2018pas8.conf...48O}. 
The transition from an accelerating to a decelerating solution happens discontinuously due to the $\abs{v'}$ dependence. The transition comes from a mismatch between the mass flux and the line force
caused either by weakening of the line force downstream or an increased mass flux. We should note that the effect of the absolute value is somewhat artificial. Physically, negative velocity gradients would provide additional resonance zones for radiation to interact with spectral lines. The additional resonance zones would create a nonlocal coupling in the gas, complicating the dynamics.

By investigating the perturbation of the line force we found that the flow is most prone to a mismatch at its base, where it is the slowest. At nearly all times and positions in the simulation, the simulations' line force perturbation and the value calculated with eq. \ref{eq:frad1Va}  match well. However, in the unsteady region nearly in HSE, the line force perturbation periodically departs from eq. \ref{eq:frad1Va}.  If instead, we calculate the gradient perturbation with 

\begin{equation}
  \label{eq:absVprime1}
  v'_1(t) = |v'(t)| - \langle v' \rangle,
\end{equation}
the line force perturbation well obeys eq. \ref{eq:frad1Va} at all positions and times. The consistency with eq. \ref{eq:frad1Va} suggests that when $v'(t)>\langle v' \rangle$, the system behaves in accordance with the linear theory. When the solution switches to a decelerating solution, $v'(t) < 0$, the nonlinear behavior is discontinuously introduced, perturbing the flow.

Because the absolute value is a non-linear operation, linear analysis does not capture the kink behavior. 
We see the departure from linear analysis in Fig. \ref{fig:Wave Sol}, occurring mainly at $r~<~1.1~r_0$. 
Additionally, we noticed that the early evolution of a solution that reaches a steady state is
very similar to that of a persistently periodic solution. Namely, during an initial transient of a steady-state solution, the innermost region produces self-excited perturbations. The main 
difference is that the perturbations gradually diminish in the steady-state solutions. 
For the periodic solutions, the evolution appeared as if it could reach a steady state if not for the process operating in the hydrostatic stratified region. 

Motivated by both factors, we reran the simulation with $r_{in}$ 
set to $1.1~r_0$ instead of $1~r_0$. 
We concurrently reduce the density using the HSE relation

\begin{equation}
    \label{eq: HSE}
   \rho(r) = \rho_0 \exp[-\hep0*(1 - r_0/r)].
\end{equation}

For this case, the initial transient exhibited a very short-lived period of
self excitation and the solution quickly attained a steady state. Thus, by moving the simulation's lower boundary outward, 
we excluded the unsteady region from the domain, preventing oscillations from being injected into the unstable region.

We ran additional simulations for various $r_{in}$ and found that the solutions are periodic
for $r_{in}<~1.1~r_0$. The periodic solutions' time-averaged density and velocity profiles 
are practically identical to the profiles of the corresponding steady solution obtained 
by moving the lower boundary outward.
We stress that we empirically determined the transition point at $r~=~1.1~r_0$ between the unsteady region where
perturbations are self-exited and the unstable region where the perturbations grow. 
We have not yet identified any clear feature in the time-averaged solution
that could be used to distinguish the two regions. For example, 
the time-averaged radial profiles presented in Figs.~\ref{fig:Decomp} and \ref{fig:CritPt} 
do not show any qualitative changes around the transition point.

Following \cite{Proga99, sundqvist2013clumping}, we also ran simulations with a reduced density and kept the location of the inner boundary constant. In the simulations, we observed that the solution evolved to a steady state. We do note however only changing the density produced solution curves that were slightly different from the solution curves of the fiducial simulations initialized with radius $r_0 = \left[G M (1 - \Gamma)\right]  \hep0^{-1} a^{-2}$ and density $\rho_0 = \mu m_p L_{\text{Edd}} \Gamma \xi_0^{-1} r_0^{-2}$. We consider these winds as originating from distinctly different objects. As such, we based our conclusions on the simulations where we modify both $r$ and $\rho$ following eq. \ref{eq: HSE}, producing a steady state and periodic solution with identical time-averaged curves.

In addition, we considered the possibility that  persistent oscillations are "leftover" from the highly
time-dependent transient from the initial conditions to the later evolutionary stage. 
Therefore, we ran simulations where we very slowly and linearly increased with time $\delta$ from $0.22$ to 0.23.
We first simulated $\delta=0.22$ solution until it reached a steady state 
and then evolved $\delta$ to a final time-dependent $\delta = 0.23$ solution. 
We observed that even the gradual increase in $\delta$ led to the emergence of self-excitation in the wind base when $\delta$ crossed a threshold between $\delta = 0.22$ and $\delta = 0.23$. The oscillations at the threshold were gentle and did not have negative velocity values (infalling material). However, as $\delta(t)$ approached $\delta = 0.23$, the velocity oscillations exhibited repetitive infalling gas.
We performed the same simulations "in reverse", meaning,  \textit{reducing} $\delta$ from $\delta = 0.23$ to $\delta = 0.22$. We found that the excitation turned off when $\delta$ approached $0.22$.

We observe two key factors that influence the self-excitation process. First, the lower boundary of the flow solution needs to be well inside the region that is almost in HSE, i.e., where the flow velocity and net force are nearly zero. Second the line force has a significant Abbott speed in the HSE region that can grow a perturbation.

Three parameters influence these factors, increasing the sensitivity to self-excitation: 
$\rho_0$, $\hep0$, and $\delta$. Increasing these three parameters makes the fluid at the inner boundary increasingly in HSE.  
In addition, $\hep0$ determines the sensitive frequencies via an increase or decrease in $\lambda_v$. While $\delta$ strongly influences the Abbott speed at the base, affecting the growth of a perturbation. 

We conjecture that the growth provided by the non-WKB behavior of the line force causes the wind to depart the steady state solution in the HSE region and to overload. Once the wind is overloaded, the non-linear behavior of $|v'|$ activates, leading to wind collapse. The wind cannot reach a steady state solution. Instead, the wind oscillates between periods of overload and infall. The resultant fluctuations develop features characteristic of acoustic waves as frequencies near the Lamb frequency continue gaining energy from the line force until they propagate out of the exponentially stratified atmosphere. 

\section{Summary and Discussion}\label{sec:Disc&Conc}

We investigated the persistent periodic behavior 
of isothermal line-driven wind solutions 
observed by \citetalias{Dannen24}. 
We identified two distinct regions in the HSE portion of the atmosphere: (i) unsteady region where perturbations are self-excited 
and (ii) unstable region where perturbations are amplified.
The unsteady region self-excites perturbations, with the fundamental frequency close
to the atmosphere's natural frequency, i.e., the Lamb cut-off frequency, see eq. \ref{eq:AcF}. The self-excitation
occurs where the time-averaged Mach number is very small (e.g., $v/a< 10^{-2}$ for the $\delta=0.3$ case).
We attribute the self-excitation to 
a mismatch between the mass flux and the line force so that
the flow undergoes repeated acceleration, deceleration, and collapse. 
The weakening of the line force due to ionization plays a significant role in 
the self-excitation behavior because the self-excitation behavior emerges 
as $\delta$ increases. The wind solution exhibits characteristic kinks in the radial velocity profiles. 
Such kinks in line-driven winds are not a new phenomenon. 
For example, \cite{feldmeier2008propagation} studied the propagation of kink behavior
by manually perturbing the density at the simulation lower boundary. 
In our work, the perturbations self-excite 
because the nearly static flow quickly decelerates to a collapsing solution
that tends to re-accelerate. As the flow re-accelerates, the line force becomes insufficient again, and the cycle repeats.

The resulting fluctuations propagate into the solution and take on the characteristics of Abbott waves. 
The line force adds energy to the waves as they travel through the remaining exponential atmosphere. 
The added energy causes the density and velocity perturbations to grow beyond what is expected 
by conserving wave energy flux, although the perturbations do not become fully non-linear.

In an exponentially stratified  atmosphere, the perturbed velocity increases exponentially 
by conserving wave energy flux \citep[e.g.,][]{clarke2007principles}. 
The exponential increase causes an intrinsic, proportional relationship between $v_1$ and $v'_1$. 
The proportional relationship allows $g_{l,1}$ to respond in phase with $v_1$,
leading to the wave energy growth. In contrast with the results of a local analysis, where $g_{l,1}$ responds $90 \degree$ out of phase 
with the acoustic perturbations, yielding a stable wave \citep{owocki2014theory}. 
As such, in our case, the wave energy growth we observe is in a non-WKB limit, i.e., the amplitude and phase change 
on a similar scale. The growth occurs dominantly for perturbations with the frequency close to $\omega_c$ 
and is confined to the exponential atmosphere.

The appearance of energy growth is sensitive to the lower boundary conditions and $\delta$. 
The location of the lower boundary and the  density, $\rho_0$, there, sets the size of the stratified atmosphere 
captured by the simulations.

In particular, it determines whether the simulation includes or excludes the unsteady region. 
The stratification is set by $\hep0$, with a large $\hep0$ indicating a colder, more stratified atmosphere. 
The line force weakening, parameterized by $\delta$,
reduces the wind velocity and velocity gradient, leading to an increase of the Abbott speed that in turn leads to instability.
The wave energy growth and self-excitation behaviors emerge as $\delta$ is a significant fraction of 
the CAK $\alpha$ parameter. 

We note that the flow properties critical to the process are not exclusive to,  $\delta>0$, mCAK models. As such, we expect much of these results to carry over to self-excited oscillations observed in $\delta = 0$ CAK winds \citep[e.g.,][]{blondin1990hydrodynamic, Proga98, Proga99b, sundqvist2013clumping}, provided there is a sufficient combination of stratification and Abbott speed in the HSE portion of the fluid.

The instability could have significant implications for X-ray binaries (XRBs) and AGNs. 
In \citetalias{Dannen24}, they showed that, for the models with the  XRB and AGN type SEDs, 
$\delta$ can be as high as $\alpha$. 
Therefore, AGN winds could be intrinsically variable due to the weakening of the line force.
This conjecture needs confirmation in 2D or 3D simulations similar to those
performed by \cite{PSK, Kurosawa09,Dyda18, Dyda24}. 

Additionally, the instability could be important for "unified" models of hot stars, such as those explored by \cite{moens2022radiation} and \cite{debnath20242d}., that seek to accurately capture the transition of a radiatively diffuse (optically thick) stellar structure to a wind in the free streaming limit. Near the sub-surface of stars, there may be acoustic instability due to radiative diffusion \cite[e.g.,][and references therein]{vandersijpt2025structureformationotypestars}. An alternate source of variability stems from the iron opacity bump in hot stars near the Eddington limit, which can drive supersonic convection in the outer envelopes \citep{2020ApJ...902...67S}. We expect, in future line-driven wind models using a Sobolev framework, that the instability identified in this work will be present. For example, the behavior could amplify manually introduced or self-excited perturbations at specific frequencies.

The self-excitation and instability, we characterized, operate in the subsonic flow 
where the Sobolev approximation is formally invalid. This approximation and the CAK line-driven
wind theory continues to be the basis of wind models because of its success in accounting for
winds in OB stars \cite[see, for example,][]{LC} and its relatively low computational cost
that makes 2D and 3D time-dependent simulations feasible. It would be important to study how the lower boundary
conditions affect the time behavior using non-Sobolev treatment of the line force such as 
full comoving-frame transfer \citep{PPK}, 
smooth source function \citep{Owocki91} 
or escape-integral source function \citep{OP96}, however simulations capturing line-deshadowing instability have been subject to this "Lamb ringing" and do overload near the base under the right circumstances \citep{sundqvist2013clumping, sundqvist20182d}.

\acknowledgments 
Support for this work was provided by the National Aeronautics and Space Administration under TCAN
grant 80NSSC21K0496. We would also like to thank Shalini Ganguly and Sergei Dyda for fruitful discussions. We thank the reviewer, Jon Sundqvist, whose comments helped us improve our presentation, particularly the presentation of the rich history of the issue of oscillations.

\newpage

\appendix

\section{Derivations}\label{sec:appendixA}
\subsection{Deriving Conservation of Acoustic Energy Equation} \label{sec:A1}

We derived the wave energy conservation equation following a similar procedure as outlined by 
 \cite{1959flme.book.....L} and \cite{mihalas1999foundations}.
By applying Eularian perturbations to the mass continuity and momentum eq. \ref{eq:ContEq} and \ref{eq:MomEq}, and using an isothermal equation of state, we obtain the perturbation equations
\begin{equation}
  \label{eq:PertMC}
  \frac{\partial \rho_{1}}{\partial t}~=~-\frac{1}{r^{2}} \frac{\partial}{\partial r} (r^{2} (v\rho_{1} + v_{1}\rho)),
\end{equation}
and 
\begin{equation}
  \label{eq:PertMom}
  \rho\frac{\partial v_{1}}{\partial t} + \rho\frac{\partial (v_{1}v)}{\partial r} + a^{2}\left(\frac{\partial \rho_{1}}{\partial r}~-~\frac{\rho_{1}}{\rho}\frac{\partial \rho}{\partial r}\right)~=~\rho g_{l,1}.
\end{equation}
By multiplying eq. \ref{eq:PertMom} by $v_{1}$ and eq. \ref{eq:PertMC} by $\frac{p_1}{\rho}$ and summing, we obtain 
\begin{equation}
  \label{eq:CAE2}
    \frac{p_1}{\rho a^{2}}\frac{\partial p_1}{\partial t} + \rho v_{1}\frac{\partial v_{1}}{\partial t} + v_{1}\left(\frac{\partial p_1}{\partial r}~-~\frac{p_1}{\rho}\frac{\partial \rho}{\partial r}\right)~=~\rho v_{1}g_{l,1}~-~ \rho v_1\frac{\partial (v_{1}v)}{\partial r}~-~\frac{p_1}{\rho}\frac{1}{r^{2}}\frac{\partial}{\partial r}\left( r^{2} \left(\rho_{1}v + \rho v_{1}\right)\right).
\end{equation}
We expand the last term on the right-hand side of eq. \ref{eq:CAE2}, cancel and regroup terms, yielding  
\begin{equation}
  \label{eq:CAEf}
  \frac{p_1}{\rho a^{2}}\frac{\partial p_1}{\partial t} + \rho v_{1}\frac{\partial v_{1}}{\partial t} + \frac{1}{r^{2}} \frac{\partial(r^{2} p_1v_{1})}{\partial r}~=~-\rho v_{1}\frac{\partial(vv_{1})}{\partial r}~-~\frac{\rho_{1}}{\rho} \frac{1}{r^{2}} \frac{\partial(r^{2} p_1v)}{\partial r} + \rho v_{1}g_{l,1}.
\end{equation}
Considering $v_1\frac{\partial v_1}{\partial t}~=~\frac{1}{2}\frac{\partial \left(v_{1}^{2}\right)}{\partial t}$, the first two terms in eq. \ref{eq:CAEf} are rewritten in the form of the conservation of acoustic energy, eq. \ref{eq:CAE}.

\subsection{Deriving Mass Flux Wave Equation} \label{sec:A2}
By following the procedure detailed in \cite{1980MNRAS.191..571P} and using equation \ref{eq:frad1Va}, we derive 
the mass flux wave equation. Specifically, we use eq. \ref{eq:pertMCf} and \ref{eq:frad1Va} to rewrite \ref{eq:PertMC} and \ref{eq:PertMom} as, 
\begin{equation}
  \label{eq:PertMC2}
  \frac{\partial \rho_{1}}{\partial t}~=~-\frac{1}{r^{2}} \frac{\partial f_1}{\partial r}, 
\end{equation}
and
\begin{equation}
  \label{eq:PertMom_2}
  \frac{\partial v_{1}}{\partial t} + \frac{\partial (v_{1}v)}{\partial r} + \frac{a^{2}}{\rho}\left(\frac{\partial \rho_{1}}{\partial r}~-~\frac{\rho_{1}}{\rho}\frac{\partial \rho}{\partial r}\right)~=~2v_{a}\left[\left(\frac{\delta}{\alpha} -1\right)\frac{\rho_{1}}{\rho}v'  + v_1'\right].
\end{equation}
We multiply eq. \ref{eq:PertMom_2} by $r^2\rho$ and substitute the relationships,
\begin{equation}
  \label{eq:T2}
    \frac{1}{\rho}\frac{\partial\rho}{\partial r}~=~\frac{-v'}{v}~-~\frac{2}{r},
\end{equation}
and
\begin{equation}
  \label{eq:T3}
    r^2\rho\frac{\partial v_{1}}{\partial t}~=~\frac{\partial f_{1}}{\partial t} + v\frac{\partial f_{1}}{\partial r} = \frac{df_1}{dt},
\end{equation}
which we obtained by using the property of mass continuity and the mass flux perturbation. The combined steps yield
\begin{equation}
  \label{eq:PertMom_f}
    \frac{\partial f_1}{\partial t} + v\frac{\partial f_1}{\partial r} + f\frac{\partial v_1}{\partial r} + v_1 r^2 \rho v' + a^2\left[r^2 \frac{\partial \rho_1}{\partial r} + r^2 \rho_1\left(\frac{v'}{v} + \frac{2}{r}\right)\right]~=~2v_a\left[\left(\frac{\delta}{\alpha}~-~1\right)v' r^2 \rho_1 + r^2 \rho \frac{\partial v_1}{\partial r}  \right].
\end{equation}
By taking the time derivative $\frac{\partial }{\partial t}$ of \ref{eq:PertMom_f}, we get 
\begin{equation}
  \label{eq:proto_Wave}
     \frac{\partial^2 f_1}{\partial t^2} + v\frac{\partial^2 f_1}{\partial r \partial t} + f\frac{\partial^2 v_1}{\partial r\partial t} + \frac{\partial v_1}{\partial t} r^2 \rho v' + a^2\left[r^2 \frac{\partial \rho_1}{\partial r\partial t} + r^2 \frac{\partial \rho_1}{\partial t}\left(\frac{v'}{v} + \frac{2}{r}\right)\right]~=~2v_a\left[\left(\frac{\delta}{\alpha}~-~1\right)v' r^2 \frac{\partial \rho_1}{\partial t} + r^2 \rho \frac{\partial v_1}{\partial r \partial t}  \right].
\end{equation}
Finally, we use eq. \ref{eq:PertMC2} to replace the $\frac{\partial\rho_1}{\partial t}$ terms and the relation \ref{eq:T3} to replace the $\frac{\partial v_1}{\partial t}$ terms. After canceling the appropriate terms, we will arrive at eq. \ref{eq:Wave} for the mass flux wave equation in an isothermal line-driven wind.

\subsection{Local Analysis of the Conservation of Energy Equation} \label{sec:A3}

Local analysis of the three terms on the RHS of eq. \ref{eq:CAE} suggests that the line force, divergence, and wave-flow interaction should remove specific energy from the wave (Key et al., submitted). For the average contribution of an acoustic cycle, the RHS of eq. \ref{eq:CAE} gives the terms
\begin{equation}
  \label{eq:Term1}
  -\left\langle\rho v_{1}\frac{\partial(vv_{1})}{\partial r} \right\rangle~=~-\rho v' \left\langle v_{1}^2\right\rangle - \rho v \left\langle v_{1}\frac{\partial v_{1}}{\partial r}\right\rangle,
\end{equation}
\begin{equation}
  \label{eq:Term2}
    -~\left\langle\frac{\rho_{1}}{\rho} \frac{1}{r^{2}} \frac{\partial(r^{2} p_1v)}{\partial r} \right\rangle = \\ -\frac{2va^2}{\rho r}\left\langle\rho_{1}^2 \right\rangle - \frac{va^2}{r}\left\langle \rho_{1}\frac{\partial \rho_1}{\partial r} \right\rangle - \frac{a^2}{\rho}v'\left\langle \rho_1^2 \right\rangle, 
\end{equation}
and 
\begin{equation}
  \label{eq:Term3}
    \left\langle \rho v_{1}g_{l,1} \right\rangle = g_l(\delta - \alpha) \left\langle \rho_1 v_1 \right\rangle + \frac{\rho g_l}{v'} \left\langle v_1 \frac{\partial v_1}{\partial r} \right\rangle.
\end{equation}

Considering a local oscillatory perturbation $q_1 \propto exp(i(kx-wt))$, quantities of the form $\langle q_1^2 \rangle$ and $\langle q_1\frac{\partial q_1}{\partial r}\rangle$ average to a positive value and zero, respectively. The term $\langle \rho_1 v_1 \rangle$ would average to a positive value for velocity and density perturbations nearly in phase. Additionally, the quantities $v', a,v,\rho,r$ are positive for a steady outflow and $\alpha>\delta$ for all cases we investigated. Collectively, this suggests the average contribution of all terms on the RHS eq. \ref{eq:CAE} is negative. We would expect the line force to remove energy from the wave.


\end{document}